\documentclass[12pt]{article}
\usepackage{a4}
\usepackage{amsmath}
\usepackage{amssymb}
\usepackage[dvips]{graphicx}
\def\beq{\begin{equation}}
\def\eeq{\end{equation}}
\def\beqa{\begin{eqnarray}}
\def\eeqa{\end{eqnarray}}
\def\n{\nonumber}

\begin{document}
 
\begin{flushright}
{SU-ITP-03/35}\\
{SAGA-HE-203}\\
{KEK-TH-924}
\end{flushright}
\vskip 0.5 truecm
 
 
\begin{center}
{\Large{\bf Ginsparg-Wilson Relation and \\
't Hooft-Polyakov Monopole on Fuzzy 2-Sphere }}\\
\vskip 1.5cm

{\large Hajime Aoki$^{a,b}$\footnote{e-mail
 address: haoki@cc.saga-u.ac.jp, haoki@itp.stanford.edu},
 Satoshi Iso$^c$\footnote{e-mail
 address: satoshi.iso@kek.jp}
 and Keiichi Nagao$^c$\footnote{e-mail
 address: nagao@post.kek.jp}}
\vskip 0.5cm
 
$^a${\it Department of Physics, Stanford University, Stanford
CA 94305-4060, USA }\\
 
$^b${\it Department of Physics, Saga University, Saga 840-8502,
Japan  }\\
 
 
$^c${\it Institute of Particle and Nuclear Studies, \\
High Energy Accelerator Research Organization (KEK)\\
Tsukuba 305-0801, Japan}
 
\end{center}
 
\vskip 1cm
\begin{center}
\begin{bf}
Abstract
\end{bf}
\end{center}
We investigate several properties of Ginsparg-Wilson
fermion on fuzzy 2-sphere.
We first examine  chiral anomaly up to the
second order of the gauge field and show that it is indeed
reduced to the correct form of the Chern character
in the commutative limit.
Next we study topologically non-trivial gauge configurations
and their topological charges.
We investigate 't Hooft-Polyakov monopole type configuration
on fuzzy 2-sphere
and show that it has the correct commutative limit.
We also consider more general configurations in
our formulation.

\newpage
\setcounter{footnote}{0}
\section{Introduction}
\setcounter{equation}{0}
 
Various matrix models have been proposed toward
nonperturbative formulations of the superstring theory.
In  matrix models like the type IIB  model \cite{IKKT},
 space-time itself is described  by matrices and thus
noncommutative (NC) geometries\cite{Connes} naturally appear
\cite{CDS,NCMM}.
Small fluctuations around the classical background
give matter degrees of freedom
and hence space-time and matter
are unified in the same matrices.
However it is not clear how space-time and matter are embedded
in matrices, for example, how metric, topology etc.
are described in matrices.
A construction of configurations with  non-trivial indices
in finite NC geometries has been an important subject
not only from the mathematical interest but also
from the physical point of view such as
the Kalza-Klein compactification of extra dimensions
with non-trivial indices to realize four dimensional
chiral gauge theories.

Topologically nontrivial configurations in finite NC
geometries have been constructed based on algebraic
K-theory and projective modules in
refs.\cite{non-trivial_config,non_chi,Valtancoli:2001gx,
Steinacker:2003sd} 
but the relations to local forms of chiral anomaly
or indices of Dirac operators are not very clear
because Dirac operators on the fuzzy sphere considered
so far
\cite{Carow-Watamura:1996wg,Grosse:1994ed,IKTW}
are not suitable to discuss these problems
in this kind of system with finite degrees of freedom.
This is summarized later in this section.
The most suitable framework will be to utilize the
Ginsparg-Wilson (GW) relation\cite{GinspargWilson} developed
in lattice gauge theory (LGT), because the GW relation
enables us to have
the exact index theorem\cite{Hasenfratzindex,Luscher} at a
finite lattice spacing using the
GW Dirac operator\cite{Neuberger}
and the modified chiral symmetry\cite{Luscher,Nieder}.

The formulation of NC geometries
in Connes' prescription is based on the spectral
triple (${\cal A}$,${\cal H}$,${\cal D}$), where a chirality
operator and a Dirac operator which anti-commute are
introduced\cite{Connes}.
In ref.\cite{AIN2}, we generalized
the algebraic relation to the GW relation
in general gauge field backgrounds
and provided prescriptions to construct
 chirality operators and  Dirac operators  which satisfy the
GW relation, so that we can define chiral structures
in finite NC geometries.
As a concrete example we considered fuzzy 2-sphere\cite{Madore}
and constructed a  set of chirality and Dirac operators.
We showed that the Chern character is obtained by evaluating
chiral anomaly up to the first order of
the gauge field\cite{AIN2}.
The evaluation of the chiral anomaly is also considered 
in refs.\cite{Ydri:2002nt,Balachandran:2003ay}. 
Fuzzy 2-sphere is one of the simplest compact NC geometry
and it can be regarded as the classical solution
of the matrix model with a Chern-Simons term\cite{IKTW}.
Since it has finite degrees of freedom and UV/IR cutoffs,
its stability can be studied analytically\cite{Imai:2003vr} or
numerically\cite{ABNN}.

We here summarize various Dirac operators on fuzzy 2-sphere.
There had been known two types of Dirac operators,
$D_{\rm WW}$\cite{Carow-Watamura:1996wg} and
$D_{\rm GKP}$\cite{Grosse:1994ed,IKTW}.
Doubling problems of these operators are studied
\cite{balagovi,balaGW} and based on this, these authors
introduced two sets of chirality operators and a free Dirac
operator satisfying GW relation.
In \cite{AIN2}, we generalized these free operators to
an interacting case in general gauge field background
configurations. Incorporation of gauge fields is
essential in the GW formalism.
Denoting this third Dirac operator as $D_{\rm GW}$,
various properties of these three types of the Dirac operators
are summarized in Table 1.
$D_{\rm WW}$ has no chiral anomaly because of the
doublers.  Both of $D_{\rm GKP}$ and $D_{\rm GW}$
have no doublers and the chiral currents
satisfy Ward identities with chiral anomaly,
whose local forms were
evaluated explicitly in ref.\cite{AIN1} and ref.\cite{AIN2}
respectively. The global form of the chiral anomaly
of $D_{\rm GKP}$ was studied in
refs.\cite{chiral_anomaly,non_chi,chiral_anomaly2}.
While in lattice gauge theories, the chiral symmetry and
the no-doubler condition have been shown to be incompatible
with some reasonable assumptions\cite{nielsen},
table 1 implies the existence of analogous no-go theorem
in finite NC geometries or matrix models.
 
\begin{table*}[htb]
\caption{The properties of three types of Dirac operators on fuzzy
 2-sphere are summarized. Each Dirac operator represents
Watamuras' operator $D_{\rm WW}$\cite{Carow-Watamura:1996wg},
Grosse et.al.'s operator $D_{\rm GKP}$\cite{Grosse:1994ed}
and Ginsparg-Wilson Dirac operator $D_{\rm GW}$.
}
\begin{center}
\renewcommand{\arraystretch}{1.1}
\begin{tabular}{|c@{\quad\vrule width0.8pt\quad}l|c|c|c|}
\hline
Dirac op. &\multicolumn{2}{c|}{chiral symmetry} & no doublers & counterpart
in LGT \\
\hline
$D_{\rm WW}$ & $D_{\rm WW} \Gamma + \Gamma D_{\rm WW} =0$ & $\bigcirc$ &
$\times$ & naive fermion \\
\hline
$D_{\rm GKP}$ & $D_{\rm GKP} \Gamma + \Gamma D_{\rm GKP} ={\cal
O}(\frac{1}{L})$ & $\times$ & $\bigcirc$ & Wilson fermion \\
\hline
$D_{\rm GW}$ & $D_{\rm GW} \hat\Gamma + \Gamma D_{\rm GW} =0$ & $\bigcirc$ &
$\bigcirc$ & GW fermion \\
\hline
\end{tabular}
\end{center}
\end{table*}

In the present paper we proceed to study the chiral
and topological properties
of the GW fermion on fuzzy 2-sphere.
After an introduction of fuzzy 2-sphere in subsection~2.1,
we provide a set of GW Dirac operator and
chirality operators on the fuzzy 2-sphere
in subsection~2.2.
We also define a topological charge and
provide an index theorem on fuzzy 2-sphere.
We prove
the index theorem in a more general formulation in appendix~A
which can be applied to any NC geometries.
The topological charge we defined takes only integer values
by definition,
and we also show in appendix~B
that its value does not change
under any small fluctuation of any parameters or
any fields in the theory.
We further need to show that it has the
correct commutative limit,
and also that it takes nonzero integer values
for topologically-nontrivial configurations.
The purpose of this paper is to
study these two points.
 
In subsection~\ref{sec:anomaly},
we evaluate chiral anomaly
by calculating the non-trivial Jacobian
under the local chiral transformation.
In NC geometries,
there is an ambiguity to define local transformations
since
the transformation parameter
do not commute with the fields and the
operators in the theory.
We here consider two chiral transformations
and corresponding chiral currents,
which transform covariantly and invariantly
under the gauge transformation respectively.
We show that
the correct form of the Chern character
is obtained for the covariant current case
by calculating the Jacobian up to the second
order in gauge fields, and taking commutative limit.
This confirms the previous result up to first order
in the gauge field in \cite{AIN2}.
Since
the topological charge we defined on the fuzzy 2-sphere
has the same form as the anomaly term,
it has the correct commutative limit.
 
In section~\ref{sec:monopole}
we study topologically nontrivial gauge configurations
and their topological charges.
Even in the theories on the commutative sphere,
after integrating the topological charge density over the sphere,
the topological charge vanishes identically for
any configurations.
For the Dirac monopole,
we need to introduce the notion of patches in the sphere
and obtain nonzero values for the topological charge.
For the 't Hooft-Polyakov (TP) monopole,
we need to introduce the idea of spontaneous symmetry breaking,
insert projector to pick up unbroken $U(1)$ component
into the topological charge,
and then obtain nonzero topological charge.
These ideas correspond to introducing some projections into
matrices in the NC theories
since both space-time and gauge group space are
embedded in matrices.
 
In subsection~3.1 we construct a TP monopole configuration
on the fuzzy sphere and show that it becomes
the correct form in the commutative limit.
The normal component of the gauge field plays the role of
the Higgs field.
We also review the topological charge for
the TP monopole in the commutative theory.
In subsection~3.2 we define a topological charge
for the TP monopole on the fuzzy 2-sphere
by introducing a projector.
In the commutative limit
this projector becomes the one to pick up the unbroken
$U(1)$ component,
and thus this topological charge has the correct
commutative limit.
In subsection~3.3 we investigate TP monopole
configurations with higher isospin.
Since it is important that these configurations
satisfy $SU(2)$ algebra,
we study its meaning.
We also define another topological charge
with the Higgs field inserted,
which has the correct commutative limit
with the projector to pick up unbroken $U(1)$ component.
We then see that the winding number of
$\pi_2 (SU(2)/U(1))$
for these configurations is $1$.
In subsection~3.4 we consider
more general configurations.
 
Similar study was done from mathematical points of view in
\cite{non-trivial_config,non_chi,Valtancoli:2001gx,Balachandran:2003ay,Steinacker:2003sd}.
In this paper
we give physical interpretation
by studying its commutative limit.
Also, we believe that
our derivation is simpler and more transparent.
 
Section~4 is devoted to conclusions and discussions.

\section{Ginsparg-Wilson fermion on fuzzy 2-sphere}
\setcounter{equation}{0}
In the paper \cite{AIN2} we proposed a general
prescription to construct  chirality operators
and Dirac operators  satisfying the
GW relation on general finite NC geometries and
gave a simple example on the fuzzy 2-sphere.
In this section
we first review this construction and calculate
the local form of anomaly up to the second order
of the gauge field.
\subsection{Brief review of fuzzy 2-sphere and $D_{\rm GKP}$}
We first briefly explain
fuzzy 2-sphere and the Dirac operator $D_{\rm GKP}$,
which will be used when we construct the GW Dirac operator $D_{\rm GW}$
in the next subsection.
 
NC coordinates of the
fuzzy 2-sphere are given by
\begin{equation}
x_i =\alpha L_i
\end{equation}
where $\alpha$ is a NC parameter,
and $L_i$'s are  $2L+1$-dimensional irreducible
representation matrices of $SU(2)$ algebra.
Thus
\begin{eqnarray}
[x_i,x_j]&=&i\alpha\epsilon_{ijk}x_k,\\
(x_i)^2&=&\alpha^2 L(L+1),
\end{eqnarray}
and
\begin{equation}
\rho=\alpha \sqrt{L(L+1)}
\end{equation}
gives the radius of the fuzzy sphere.
 
Wave functions on fuzzy 2-sphere are composed of
$(2L+1)\times(2L+1)$ matrices, and can be expanded
in terms of NC
analogues of the spherical harmonics $\hat{Y}_{lm}$.
They are traceless symmetric products of
the NC coordinates.
There is an upper bound for the angular momentum
$l$ in $\hat{Y}_{l,m}$:
$l \le 2L$.
Derivatives along the
Killing vectors on the sphere are given by
the adjoint action of $L_i$ as
\begin{equation}
{\cal L}_i M= [L_i, M] =(L_i^L-L_i^R)M,
\end{equation}
where $M$ is any wavefunction and
the superscript L (R) in $L_i$ means that this
operator acts from the left (right) on matrices.
An integral over the 2-sphere is expressed by a trace
over matrices:
\begin{equation}
\frac{1}{2L+1} {\text tr} \leftrightarrow
\int \frac{d \Omega}{4 \pi}.
\end{equation}
 
The fermionic action with the Dirac operator $D_{\rm GKP}$
is given by\cite{Grosse:1994ed,IKTW}
\begin{eqnarray}
&&S_{\rm GKP}={\rm tr}(\bar\psi D_{\rm GKP} \psi), \\
&&D_{\rm GKP}=\sigma_i({\cal L}_i+\rho a_i)+1.
\end{eqnarray}
Here $i$ runs from $1$ to $3$,
$\sigma_i$'s are Pauli matrices,
and
the fermion field $\psi$ and the gauge field $a_i$ are expressed
as $(2L+1)\times(2L+1)$ Hermitian matrices.
The free part of this theory does not have doubler
modes \cite{balagovi, AIN1}.
 
$S_{\rm GKP}$ is invariant
under the following gauge transformation:
\begin{eqnarray}
\psi &\rightarrow& U\psi, \quad
\bar\psi \rightarrow \bar\psi U^\dag, \label{gaugeTrfer}\\
a_i &\rightarrow& U a_i U^\dag +\frac{1}{\rho} (U L_i U^\dag-L_i),
\label{gaugeTrai}
\end{eqnarray}
since a combination
\begin{equation}
A_i \equiv L_i+\rho a_i
\label{defAi}
\end{equation}
transforms covariantly under the gauge transformation as
\begin{equation}
A_i \rightarrow U A_i U^\dagger, \label{gaugetr_A}
\end{equation}
and the fermion $\psi$ transforms as  the fundamental
representation.
 
The normal component of $a_i$ to the sphere can
be interpreted as
the scalar field on the sphere.
We define it covariantly as
\begin{equation}
\phi=\frac{A_i^2 -L(L+1)}{2L+1} \frac{1}{\rho}
=\frac{\phi'}{\rho}\label{def_NC_phi}.
\end{equation}
Here we also defined a normalized scalar field
$\phi'$ for later convenience.
 
The commutative limit can be taken by
$\alpha \to0, L \to \infty$
with $\rho$ fixed.
Then,
$D_{\rm GKP}$
becomes a Dirac operator on the commutative 2-sphere,
\begin{equation}
D_{\rm com} = \sigma_i ({\tilde {\cal L}_i}+\rho a_i) +1,
\end{equation}
where
\begin{equation}
{\tilde {\cal L}_i}=-i\epsilon_{ijk}x_j \partial_k.
\label{comLi}
\end{equation}
The scalar field (\ref{def_NC_phi})
becomes the one on the commutative 2-sphere,
$a_i n_i$,
where
$n_i=x_i/\rho$.
 
More detailed explanations are given
in refs.\cite{IKTW,AIN1,AIN2}.

\subsection{Ginsparg-Wilson fermion on fuzzy 2-sphere}
In this subsection we review the construction of
the GW Dirac operator
\cite{AIN2}.
 
We first introduce two hermitian chirality operators
$\Gamma^R$ and $\hat\Gamma$ satisfying
$\hat\Gamma^2= (\Gamma^R)^2=1$ as follows:
\begin{eqnarray}
&&\Gamma^R = a\left(\sigma_i L_i^R -\frac{1}{2}\right)
=\frac{\sigma_i L_i^R -\frac{1}{2}}
{\sqrt{(\sigma_i L_i^R -\frac{1}{2})^2}}, \label{gammaR}\\
&&\hat\Gamma=\frac{H}{\sqrt{H^2}} \label{gammahat},
\end{eqnarray}
where
\begin{equation}
a=\frac{1}{L+\frac{1}{2}}
\end{equation}
is introduced as a NC analogue of a lattice-spacing.
The superscript $R$ of $L_i^R$ means that this operator
acts from the right on matrices.
We define the Hermitian operator $H$ as
\begin{eqnarray}
H&=&a\left(\sigma_i A_i +\frac{1}{2}\right)\\
&=&\Gamma^R + a D_{\rm GKP}\\
&=& \Gamma^L +a\rho\sigma_i a_i^L, \label{L}
\end{eqnarray}
where
\begin{equation}
\Gamma^L=a\left(\sigma_i L_i^L +\frac{1}{2}\right)
=\frac{\sigma_i L^L_i +\frac{1}{2}}
{\sqrt{(\sigma_i L^L_i +\frac{1}{2})^2}}, \label{gammaL}
\end{equation}
is the Hermitian chirality operator
introduced in \cite{balaGW} and satisfies
$(\Gamma^L)^2=1$.
In eqs.(\ref{L}),(\ref{gammaL}) the superscript
$L$ in $a_i^L$ and $L_i^L$ means that this
operator acts from the left on matrices.
 
We thus see that $\hat\Gamma$ in eq.(\ref{gammahat})
becomes $\Gamma^L$ for vanishing gauge fields.
We also note that both of the two chirality operators
$\Gamma^R$ and $\Gamma^L$ become the
chirality operator $\gamma=\sigma_i n_i$
in the commutative limit.
This was discussed in \cite{balaGW} for the free case
without gauge field backgrounds.
 
We next define
the GW Dirac operator as
\begin{equation}
D_{\rm GW} = -a^{-1}\Gamma^R (1- \Gamma^R \hat{\Gamma}).
\label{defDGW}
\end{equation}
The fermionic action
\begin{equation}
S_{\rm GW}={\text tr}(\bar{\Psi} D_{\rm GW} \Psi) \label{SGW}
\end{equation}
is invariant under the gauge transformation
(\ref{gaugeTrfer}), (\ref{gaugetr_A})
because $D_{\rm GW}$ is composed of $\Gamma^R$ and $A_i$.
The free part of the theory has no doublers.
In the commutative limit $D_{\rm GW}$ becomes the
commutative Dirac operator without
coupling to the scalar field $\phi$,
\begin{eqnarray}
D_{\rm GW}&\simeq& (D_{\rm GKP}-\{\Gamma^R,D_{\rm GKP} \}\Gamma^R/2)
+{\cal O}(1/L) \n \\
& \to & \sigma_i (\tilde{\cal L}_i + \rho P_{ij} a_j) +1 ,
\end{eqnarray}
where $P_{ij}$ is a projection operator on the sphere,
\begin{equation}
P_{ij}=\delta_{ij}-n_i n_j.
\end{equation}
This satisfies $(P^2)_{ij} = P_{ij}$ and $n_i P_{ij} =0$.
 
We can see from the definition (\ref{defDGW})
that $D_{\rm GW}$ satisfies the GW relation:
\begin{equation}
\Gamma^R D_{\rm GW}+D_{\rm GW} \hat{\Gamma}=0.\label{GWrelation}
\end{equation}
Then, as we show in appendix \ref{sec:indextheo},
we can prove the following index theorem:
\begin{equation}
{\rm{index}}D_{\rm GW}\equiv (n_+ - n_-)=\frac{1}{2}
{\cal T}r(\Gamma^R +\hat{\Gamma}), \label{indexth}
\end{equation}
where
$n_{\pm}$ are the numbers of zero eigenstates of $D_{\rm GW}$
with a positive (or negative) chirality (for either $\Gamma^R$
or $\hat{\Gamma}$)
and ${\cal T}r$ is a trace of operators acting on matrices.
We also prove in appendix \ref{sec:deltrhatgamma}
that ${\cal T}r (\hat{\Gamma})$ is invariant
under a small deformation of any parameter or any
configuration such as gauge field in the operator $H$.
Furthermore, $\frac{1}{2}{\cal T}r(\Gamma^R +\hat{\Gamma})$
takes only integer values since both $\Gamma^R$ and $\hat{\Gamma}$
have a form of sign operator
by the definitions (\ref{gammaR}) (\ref{gammahat}).
Therefore, we may call this a topological charge.
 
In the next subsection we will investigate the commutative limit
of this topological charge,
and show that it becomes the Chern character
in the commutative limit.
In section~\ref{sec:monopole}
we will investigate topologically nontrivial
configurations and their topological charges.

\subsection{Chiral anomaly}\label{sec:anomaly}
In this subsection we calculate the local form of the chiral anomaly.
 
The fermionic action (\ref{SGW}) is invariant
under the global chiral
transformation,
\begin{equation}
\delta\Psi=i \hat{\Gamma}\Psi,\
\delta\bar\Psi=i \bar\Psi \Gamma^R,
\end{equation}
due to the GW relation (\ref{GWrelation}).
For a local transformation, however, we need to specify the
ordering of the chiral transformation parameter $\lambda$,
the fermion field, and the chirality operator,
since they are not commutable.
This ambiguity is specific to the NC field theories
and makes the analysis of the Ward-Takahashi(WT)
identity complicated\cite{anomalyNC,AIN1,AIN2}
 
Here we consider two types of local chiral transformations.
The first type of chiral transformation is defined as
\begin{equation}
 \delta \Psi =i \lambda \hat{\Gamma} \Psi, \ \
 \delta \bar{\Psi} = i \bar{\Psi}\lambda \Gamma^R,
\label{covchiraltr}
\end{equation}
where the chiral transformation parameter $\lambda$
should transform covariantly as
$\lambda \rightarrow U \lambda U^\dag$
under the gauge transformation
(\ref{gaugeTrfer}),(\ref{gaugeTrai}).
The associated chiral current transforms covariantly.
Another chiral transformation is defined as
\begin{equation}
 \delta \Psi =i  \hat{\Gamma} \Psi \lambda, \ \
 \delta \bar{\Psi} = i \lambda \bar{\Psi} \Gamma^R,
\label{invchiraltr}
\end{equation}
where the chiral transformation parameter $\lambda$
is assumed to be invariant under
gauge transformations,
so is the associated chiral current.
 
In the WT identity for the gauge-covariant current,
the variation of the action (\ref{SGW}) under (\ref{covchiraltr})
gives the
current-divergence term,
and the variation of the integration measure gives the anomaly term,
\begin{eqnarray}
2q_{\rm cov}(\lambda)&\equiv&
{\cal T}r
(\lambda^L \hat{\Gamma} +\lambda^L \Gamma^R) \n \\
&=&{\text tr}(1)
{\text Tr} (\lambda \hat{\Gamma})
+ {\text tr}(\lambda) {\text Tr}(\Gamma^{R})\n \\
&=&\frac{2}{a}
{\text Tr} (\lambda \hat{\Gamma})
-2{\text tr}(\lambda), \label{covtcd}
\end{eqnarray}
where in the first line,
${\cal T}r$ is a trace of operators acting on matrices,
and the superscript $L$ ($R$) means that this
operator acts from the left (right) on matrices.
In the second and third lines,
${\text Tr}$ is a trace over matrices and spinors,
${\text tr}$ is a trace over matrices,
and $\Gamma^R$ and $\hat{\Gamma}$
are considered as mere matrices instead of operators
acting on matrices.
Similarly, in the WT identity for the gauge-invariant current,
the variation of the measure under (\ref{invchiraltr})
gives the anomaly term,
\begin{eqnarray}
2q_{\rm inv}(\lambda)&\equiv&
{\cal T}r
(\lambda^R \hat{\Gamma} +\lambda^R \Gamma^R) \n \\
&=&{\text tr}(\lambda)
{\text Tr} (\hat{\Gamma})
+ {\text tr}(1) {\text Tr}(\lambda \Gamma^{R})\n \\
&=&{\text tr}(\lambda)
{\text Tr} (\hat{\Gamma})
- 2{\text tr}(\lambda).\label{invtcd}
\end{eqnarray}
For a global chiral transformation, that is, $\lambda=1$ case,
both $q_{\rm cov}(\lambda)$ and $q_{\rm inv}(\lambda)$
become the topological charge defined in eq.(\ref{indexth}).
When background gauge fields vanish, $\hat{\Gamma}=\Gamma^L$,
and  $q_{\rm cov}(\lambda)$ and $q_{\rm inv}(\lambda)$ vanish.
 
For the covariant case (\ref{covtcd}), the chiral transformation
parameter $\lambda$ and the gauge field $a_i$ in $\hat\Gamma$
are inserted in the same trace,
while for the invariant case (\ref{invtcd}),
$\lambda$ and $a_i$ are inserted in different traces.
Since traces are replaced by the integrations in the
commutative limit, this indicates that local WT identity can
be written down for the covariant current, while some
nonlocality must be introduced in the WT identity for
the invariant current.
This is consistent with the previous results
\cite{anomalyNC, AIN1}.
Gauge invariant operators can be defined only by taking
traces over matrices,
and thus this introduces nonlocality in NC spaces.
 
We now consider weak gauge field configurations,
and see if the topological charge density $q_{\rm cov}(\lambda)$
reduces to
the Chern character in the commutative limit.
Expanding the topological charge density (\ref{covtcd})
with respect to the gauge field $a_i$
up to the second order, and
taking a trace over $\sigma$ matrices, we obtain
\begin{eqnarray}
q_{\rm cov}(\lambda)&=&
\frac{a^2 \rho^2}{\alpha}i \
{\text tr} \Bigl( \lambda [L_i, a'_i] \Bigr) \n \\
&&+{\text tr} \biggl[
\lambda \biggl(
\frac{3}{8}a^4 \rho^2
\Bigl\{ [L_i,a_i]^2 - 4\left(\frac{\rho}{\alpha}\right)^2 (a'_i)^2
+4i \frac{\rho}{\alpha} L_i \{[L_j,a_j],a'_i\} \n \\
&&-8i \left(\frac{\rho}{\alpha}\right)^2
\epsilon_{ijk}L_i a'_j a'_k \Bigr\}
-a^2 \rho^2 i \frac{\rho}{\alpha} [a_i,a'_i]
\biggr) \biggr] \n \\
&&+{\cal{O}}(a_i^{\prime 3}), \label{2ndorder}
\end{eqnarray}
where
\begin{equation}
a'_i=\frac{\alpha}{2 \rho}\epsilon_{ijk}(L_j a_k + a_k L_j)
\end{equation}
is the tangential component of
the gauge field $a_i$. The gauge field $a_i$ are decomposed
into the tangential component $a'_i$ and the
normal component, that is, the scalar field $\phi$.
 
In the commutative limit,
the second line of (\ref{2ndorder}) vanishes, and we obtain
\begin{equation}
q_{\rm cov}(\lambda) \to \rho^2 \int  \frac{d \Omega}{4 \pi}
{\text tr}\Big[
\lambda \epsilon_{ijk} \frac{x_i}{\rho} F_{jk} \Big],
\label{chern}
\end{equation}
where the trace is taken over the non-Abelian gauge group,
$F_{jk}$ is the field strength defined as
$F_{jk}= \partial_j a_k'-\partial_k a_j'-i[a_j',a_k']$,
and $a'_i = {\epsilon_{ijk}x_j a_k / \rho }$.
For the $U(1)$ gauge theory,
the trace in (\ref{chern}) is not necessary
and the field strength is defined as such.
All contributions from higher order terms in the
gauge field $a_i$
vanish in the commutative limit.
It is now confirmed that the Chern character is reproduced
in all orders in $a_i$, which was previously
checked up to the first order of $a_i$ in \cite{AIN2}.
This topological charge density is nothing but the magnetic
flux density penetrating the 2-sphere,
and we obtain the correct form of anomalous WT identity
in the commutative limit.
Since the topological charge we defined in (\ref{indexth})
has the form of $q_{\rm cov}(\lambda)$ with $\lambda=1$
in (\ref{covtcd}),
our topological charge (\ref{indexth})
becomes the Chern character
in the commutative limit,
at least locally.
 
However, for $\lambda=1$,
even in the commutative theories,
(\ref{chern}) vanishes identically
for any configurations
after the integration over the sphere,
if the gauge field $a_i$ is a single-valued function
on the sphere
for Abelian gauge theory,
or
if we naively take the trace for non-Abelian gauge theory.
In order to describe topologically-nontrivial configurations
and classify them by some topological charge,
we need to introduce such ideas as
patches in the Dirac monopole, or spontaneous symmetry breakings in
the TP monopole.
In NC theories,
these ideas correspond to introducing some kind of
projections into matrices,
since both space-time and gauge group space are
embedded in matrices in NC theory.
We will study this subject in the
next section.

\section{'t Hooft-Polyakov monopole on fuzzy 2-sphere}
\label{sec:monopole}
\setcounter{equation}{0}
 
In this section we construct topologically nontrivial
configurations which correspond to
the 't Hooft-Polyakov (TP) monopole in the commutative theory.
We then define a topological charge
for nontrivial configurations by
introducing projection operators.
We show that the topological charge has
the correct commutative limit.
Similar study was done by using projective modules
\cite{non-trivial_config,non_chi,Valtancoli:2001gx,
Balachandran:2003ay,Steinacker:2003sd}.
In the following we will give physical interpretation
of the previous mathematical settings
by studying its commutative limit.
 
\subsection{TP monopole configuration on fuzzy 2-sphere}
\label{sec:tpmpconf}
Let's consider the following configuration
in the $SU(2)$ gauge theory on fuzzy 2-sphere:
\begin{equation}
a_i=\frac{1}{\rho}{\bold 1}_{2L+1} \otimes \frac{\tau_i}{2},
\label{monopoleconfig}
\end{equation}
where the first and second factors represent
the NC space and the gauge group space
respectively.
Then, the combination (\ref{defAi}) becomes
\begin{equation}
A_i= L_i^L \otimes {\bold 1}_2 +
{\bold 1}_{2L+1} \otimes \frac{\tau_i}{2}.\label{AiLtau}
\end{equation}
We note here that $A_i$'s satisfy the $SU(2)$ algebra:
\begin{equation}
[A_i, A_j]=i\epsilon_{ijk}A_k.
\label{SU2alg}
\end{equation}
We will use this property for constructing a nontrivial
topological charge
on the fuzzy sphere in the next subsection.
Here we will first show that
the configuration (\ref{monopoleconfig})
corresponds to the section of
the TP monopole configuration on $S^2$
in the commutative theory.
We will also see that the scalar field $\phi$,
the normal component of the gauge field $a_i$,
has a role of the Higgs field in
the $SU(2)$ adjoint representation.
 
In the commutative limit, (\ref{monopoleconfig}) becomes
\begin{equation}
a_i(x)=\frac{1}{\rho}\frac{\tau_i}{2},\label{gauge_configuration}
\end{equation}
or
\begin{equation}
a_i^a(x)=\frac{1}{\rho}\delta_{ia}, \label{configuration}
\end{equation}
if decomposed by
$a_i=a_i^a \tau^a/2$.
We further decompose it into the tangential component
on 2-sphere, $a'_i$ and normal component $\phi$  by
\begin{eqnarray}
&&\left\{
\begin{array}{lll}
a_i'&=& \epsilon_{ijk}n_j a_k, \\
\phi&=&n_i a_i,
\end{array}
\right. \label{decomposeto}\\
&&a_i = -\epsilon_{ijk}n_j a_k' + n_i \phi,
\label{decomposefrom}
\end{eqnarray}
where $n_i=x_i/\rho$. Then we obtain
\begin{eqnarray}
a'^a_i&=&\frac{1}{\rho^2}\epsilon_{ija} x_j,
\label{TPsolutionai}\\
\phi^a&=&\frac{1}{\rho}n_a,
\label{TPsolution}
\end{eqnarray}
for the configuration (\ref{configuration}).
These are nothing but the TP monopole
configurations.

This configuration satisfies
\begin{equation}
a'^a_i=-\epsilon^{abc}\phi'^b \partial_i \phi'^c,
\label{puregaugeconfig}
\end{equation}
where
\begin{equation}
\phi'^a=\frac{\phi^a}{\sqrt{(\phi^a)^2}}=n_a
\label{HHphiprime}
\end{equation}
is the normalized scalar field which satisfy
$\sum_a(\phi'^a)^2=1$.
For (\ref{puregaugeconfig}),
the covariant derivative of $\phi'$ and
the field strength become
\begin{eqnarray}
(D_i \phi')^a &\equiv &\partial_i \phi'^a+\epsilon^{abc}a_i'^b \phi'^c\\
&=&0, \\
F_{ij}^a &\equiv& \partial_i a_j'^a -\partial_j a_i'^a
+\epsilon^{abc} a_i'^b a_j'^c \\
&=& -2\epsilon^{abc} (\partial_i \phi'^b) (\partial_j \phi'^c)
+\epsilon^{bcd}\phi'^a \phi'^b(\partial_i \phi'^c)(\partial_j \phi'^d).
\label{fieldstrengthpure}
\end{eqnarray}
If we extend this configuration to 3-dimensional space,
by regarding $\rho$ as the radial coordinate,
we obtain the following asymptotic behavior:
\begin{equation}
\left\{
\begin{array}{l}
(\phi'_a)^2 \rightarrow 1, \\
D_i \phi' \rightarrow 0,\\
F_{ij} \rightarrow O(1/\rho^2),
\end{array}
\right.
{\rm for} \quad \rho  \rightarrow \infty,  \label{TPcondition}
\end{equation}
which assures the finiteness of the energy defined in
3-dimensions.
 
We next consider the winding number of $\pi_2 (SU(2)/U(1))$.
The magnetic flux of unbroken $U(1)$ component penetrating the
2-sphere is written as
\begin{eqnarray}
Q
&=&\frac{\rho^2}{4\pi}\int_{S^2} d\Omega
{\text tr}(P_\tau \epsilon_{ijk}
n_i F_{jk})  \label{comTIP}\\
&=& \frac{\rho^2}{8\pi}\int_{S^2} d\Omega \epsilon_{ijk}
n_i \phi'^a F_{jk}^a,\label{comTIphi}
\end{eqnarray}
where $P_\tau=\frac{1+\tau_i n_i}{2}$ is the projector
to pick up the unbroken $U(1)$ component.
We note here that this is the Chern character (\ref{chern})
with the projection operator $P_{\tau}$ or $\phi'$ inserted.
In the configuration (\ref{puregaugeconfig}),
using (\ref{fieldstrengthpure}),
we obtain
\begin{equation}
Q=-\frac{\rho^2}{8\pi}\int_{S^2} d\Omega \epsilon_{ijk}
n_i \epsilon^{abc} \phi'^a
(\partial_j \phi'^b) (\partial_k \phi'^c).
\end{equation}
Thus $-Q$ is the degree of mapping of
$\phi'^a(x_i)$ : $S_x^2 \mapsto S_\phi^2$, $\pi_2(S^2)={\bold Z}$.
Inserting (\ref{HHphiprime}), this gives
\begin{equation}
Q=-\frac{1}{4\pi}\int_{S2} d\Omega =-1 \label{com_index}.
\end{equation}
Thus the configuration (\ref{monopoleconfig})
corresponds to the TP monopole configuration
in the commutative theory.

\subsection{Topological charge of TP monopole on fuzzy 2-sphere}
 
In this subsection we study the topological charge
for the configuration (\ref{monopoleconfig})
by introducing a projection operator into the
topological charge in (\ref{indexth}).
We see that this topological charge
corresponds to that in commutative theory
which was studied at the end of the previous subsection.

Since $A_i$'s in (\ref{AiLtau}) satisfy $SU(2)$ algebra,
we can decompose $A_i$ into irreducible representations
using some unitary matrix $U$ as
\begin{equation}
A_i= U
\begin{pmatrix}
 L_i^{(1)} & \cr
& L_i^{(2)} \cr
\end{pmatrix}
U^\dag \label{decomposition},
\end{equation}
where $L_i^{(1)}$ and $L_i^{(2)}$ are
$L^{(1)}=L+\frac{1}{2}$ and $L^{(2)}=L-\frac{1}{2}$
representations respectively.
We denote the Hilbert spaces on which
the operator $L^{(1)}_i$ and $L^{(2)}_i$ act
as ${\cal H}^{(1)}$ and ${\cal H}^{(2)}$
respectively.
Each Hilbert space can be picked up by the following
projection operators:
\begin{eqnarray}
P^{(1)}
&=& \frac{(A_i)^2-L^{(2)}(L^{(2)}+1)}
{L^{(1)}(L^{(1)}+1)-L^{(2)}(L^{(2)}+1)} \label{P1Ai}\\
&=&\frac{1+\Gamma_\tau}{2} \label{P1tau} \\
&=& \rho\phi + \frac{L +\frac{1}{4}}{2L+1},\label{P1phi}\\
P^{(2)}
&=& \frac{L^{(1)}(L^{(1)}+1)-(A_i)^2}
{L^{(1)}(L^{(1)}+1)-L^{(2)}(L^{(2)}+1)} \label{P2Ai}\\
&=&\frac{1-\Gamma_\tau}{2} \label{P2tau}\\
&=&-\rho\phi + \frac{L +\frac{3}{4}}{2L+1} \label{P2phi}
\end{eqnarray}
where the Hermitian operator
\begin{equation}
\Gamma_\tau=\frac{L_i \tau_i+\frac{1}{2}}{L+\frac{1}{2}}
\end{equation}
satisfies $(\Gamma_\tau)^2=1$ and becomes
$\tau_i n_i$ in the commutative limit.
The scalar field $\phi$ in (\ref{P1phi}) (\ref{P2phi})
is defined in (\ref{def_NC_phi}).
 
We define the chirality operator $\hat\Gamma$
as in (\ref{gammahat}).
This has the following block diagonal form
when $A_i$ are decomposed into
irreducible representations:
\begin{equation}
\hat\Gamma= U
\begin{pmatrix}
 \hat\Gamma^{(1)} & \cr
& \hat\Gamma^{(2)} \cr
\end{pmatrix}
U^\dag \label{gammadecomposition},
\end{equation}
where
\begin{equation}
\hat\Gamma^{(a)} = \frac{\sigma_i L_i^{(a)} + \frac{1}{2}}
{L^{(a)} + \frac{1}{2} },
\end{equation}
for $a=1,2$.
 
We now consider the topological charge (\ref{indexth})
with the projector (\ref{P1Ai})-(\ref{P2phi})
\begin{equation}
\frac{1}{2}{\cal T}r P^{(a)} (\Gamma^R + \hat\Gamma)=
\frac{1}{2}{\cal T}r_{(a)}( \Gamma^R + \hat\Gamma^{(a)} ),
\label{proj_index}
\end{equation}
where the trace ${\cal T}r$ is taken over the whole Hilbert space
on which
$A_i$, $-L_i^R$, $\frac{\sigma_i}{2}$ operate.
The trace ${\cal T}r_{(a)}$ is taken over the restricted
Hilbert space ${\cal H}^{(a)}$.
We note that, since $A_i$'s satisfy the $SU(2)$ algebra,
the projector $P^{(a)}$ commutes with
$\Gamma^R$ and $\hat\Gamma$.
We can see that the topological charge (\ref{proj_index})
is invariant under any small perturbations
and take only integer values
since this has the form of a sign operator
as we mentioned below (\ref{indexth})
\footnote{
We can also define a GW Dirac operator as (\ref{defDGW})
multiplied by the projection operator $P^{(a)}$,
and show that the index theorem is satisfied
between the index for this Dirac operator
and the topological charge
(\ref{proj_index}).
However, the meaning of this Dirac operator is not
clear.
}.
If we take the weak gauge limit and the commutative limit
of (\ref{proj_index}) as we did in (\ref{chern}),
we obtain
\begin{equation}
\frac{1}{2}{\cal T}r P^{(a)} (\Gamma^R + \hat\Gamma)
\to \rho^2 \int  \frac{d \Omega}{4 \pi}
{\text tr}\Big[
P^{(a)} \epsilon_{ijk} n_i F_{jk} \Big].
\label{comlimPTC}
\end{equation}
By using (\ref{P1tau})(\ref{P1phi}) for $P^{(1)}$,
this becomes (\ref{comTIP})(\ref{comTIphi}),
while
by using (\ref{P2phi}) for $P^{(2)}$,
this becomes a product of $-1$ and (\ref{comTIphi}).
We thus see that the topological charge (\ref{proj_index})
becomes the desired form in the commutative limit.

We will now evaluate the toplogical charge (\ref{proj_index}).
It was done in ref.\cite{Balachandran:2003ay},
but we will give simpler evaluation below.
For evaluating the first term,
$\frac{1}{2}{\cal T}r_{(a)} \Gamma^R$,
we introduce an operator $J_i=\frac{\sigma_i}{2}-L_i^R$.
The Casimir operator of $J_i$ is given by
\begin{equation}
(J_i)^2=J(J+1)=L(L+1)+\frac{3}{4}-\sigma_i L_i^R.
\end{equation}
Denoting the degeneracy as $m$,
we obtain
\begin{itemize}
\item For $J= L+\frac{1}{2}$, \quad $\Gamma^R=-1$,\quad
$m=(2L+2)(2L^{(a)}+1)$,
\item For $J= L-\frac{1}{2}$, \quad $\Gamma^R=1$,\quad
$m=2L(2L^{(a)}+1)$,
\end{itemize}
and thus
\begin{equation}
\frac{1}{2}{\cal T}r_{(a)} \Gamma^R =-(2L^{(a)}+1).\label{trGamma}
\end{equation}
 
Similarly, for evaluating the second term,
$\frac{1}{2}{\cal T}r_{(a)} \hat\Gamma^{(a)}$,
we introduce another operator
$J_i'=\frac{\sigma_i}{2}+L_i^{(a)}$.
The Casimir operator of $J_i'$ is calculated as
\begin{equation}
(J_i')^2=J'(J'+1)=L^{(a)}(L^{(a)}+1)
+\frac{3}{4} + \sigma_i L_i^{(a)}.
\end{equation}
Then we obtain
\begin{itemize}
\item For $J'= L^{(a)}+\frac{1}{2}$,
\quad $\hat\Gamma^{(a)}=1$, \quad
$m=(2L^{(a)}+2)(2L+1)$,
\item For $J'= L^{(a)}-\frac{1}{2}$,
\quad $\hat\Gamma^{(a)}=-1$, \quad
$m=2L^{(a)}(2L+1)$,
\end{itemize}
and
\begin{equation}
\frac{1}{2}{\cal T}r_{(a)} \hat\Gamma^{(a)} = 2L+1.
\label{trHatGamma}
\end{equation}

{}From eqs.(\ref{trGamma})(\ref{trHatGamma}), we obtain
\begin{equation}
\frac{1}{2}{\cal T}r P^{(a)} (\Gamma^R + \hat\Gamma)
=\frac{1}{2}{\cal T}r_{(a)}( \Gamma^R + \hat\Gamma^{(a)} )
= 2(L-L^{(a)}) . \label{NCindex}
\end{equation}
For $a=1,2$, $L^{(1)}=L+1/2, L^{(2)}=L-1/2$, and
\begin{eqnarray}
\frac{1}{2}{\cal T}r P^{(1)} (\Gamma^R + \hat\Gamma)
 &=& \frac{1}{2} {\cal T}r
\left(\rho\phi (\Gamma^R + \hat\Gamma) \right)=-1, \label{PTCfor1}\\
\frac{1}{2}{\cal T}r P^{(2)} (\Gamma^R + \hat\Gamma)
&=& -\frac{1}{2}
{\cal T}r \left(\rho\phi (\Gamma^R + \hat\Gamma) \right)=1.
\label{PTCfor2}
\end{eqnarray}
This result agrees with
the monopole charge $Q$ in the commutative theory
which was calculated
in (\ref{com_index}) in the previous subsection.
We have thus shown that the non-trivial configuration
(\ref{monopoleconfig})
can be
interpreted as the TP monopole configuration,
and the topological charge (\ref{proj_index})
for this configuration
gives the
correct value.

\subsection{Higher isospin}
\label{sec:higherisospin}
 
In this subsection we consider a configuration
coupled to a fermion with higher
isospin $T$\cite{non-trivial_config,non_chi,Balachandran:2003ay}:
\begin{equation}
a_i=\frac{1}{\rho}{\bold 1}_{2L+1} \otimes T_i,
\label{higerisospinai}
\end{equation}
where $T_i$'s are the $2T+1$ dimensional
representation of $SU(2)$
algebra.
Since the combination
\begin{equation}
A_i=L_i^L \otimes {\bold 1}_{2T+1} +
{\bold 1}_{2L+1} \otimes T_i
\label{higherisospinconfig}
\end{equation}
satisfies $SU(2)$ algebra,
it can be decomposed into the irreducible representations
of $SU(2)$ as
\begin{equation}
A_i \simeq
\begin{pmatrix}
 L_i^{(1)} & & & \cr
& L_i^{(2)} & & \cr
& & \ddots & \cr
& & & L_i^{(2T+1)} \cr
\end{pmatrix}
\end{equation}
where
$L_i^{(a)}$'s ($a=1,\cdots,2T+1 $) denote the $L^{(a)}=L+T+1-a$
representations respectively.
 
Projectors to pick up each Hilbert space on which
$L_i^{(a)}$ acts can be
defined as in (\ref{P1Ai})(\ref{P2Ai}),
\begin{equation}
P^{(a)}=\prod_{b\neq a}
\frac{(A_i)^2 -L^{(b)}(L^{(b)}+1)}{L^{(a)}(L^{(a)}+1)-L^{(b)}(L^{(b)}+1)}.
\label{Pamulti}
\end{equation}
Then we define the same topological charge
as (\ref{proj_index}),
and obtain the same result as (\ref{NCindex}).

Both of the configurations (\ref{AiLtau})
(\ref{higherisospinconfig}) satisfy the $SU(2)$ algebra
(\ref{SU2alg}).
Since it is essential in the calculations of
the topological charge (\ref{proj_index}),
we here study its meaning.
By inserting (\ref{defAi}) into (\ref{SU2alg}),
taking the commutative limit,
using (\ref{comLi}),
and decomposing $a_i$ into $a'_i$ and $\phi$
by (\ref{decomposefrom}),
we obtain
\begin{equation}
\epsilon_{ilm} \epsilon_{jpq} n_l n_p F_{mq}
+ (n_i \epsilon_{jlm} - n_j \epsilon_{ilm}) n_l (D_m \phi)
+\frac{1}{\rho} \epsilon_{ijk} n_k \phi
=0.
\label{SU2com}
\end{equation}
Since  (\ref{SU2com}) is a second rank antisymmetric tensor
with indices $i,j$,
we can decompose  it into
normal-tangential and tangential-tangential components on
the sphere,
and obtain
\begin{eqnarray}
\epsilon_{ijk} n_j (D_k \phi) &=&0, \label{SU2com1}\\
\epsilon_{ijk} n_i F_{jk} + \frac{2}{\rho} \phi &=& 0.\label{SU2com2}
\end{eqnarray}
Eq. (\ref{SU2com1}) means that the tangential component
of $D_i \phi$ vanishes,
while (\ref{SU2com2}) means that the normal component of the
flux is equal to the scalar field.
We see that the finite energy condition (\ref{TPcondition})
is satisfied by (\ref{SU2com1})(\ref{SU2com2})
since $\phi=\phi'/\rho$ and $(\phi'_a)^2=1$.
Also, by (\ref{SU2com2})
the topological charge (\ref{comlimPTC})
becomes
\begin{equation}
-2 \int  \frac{d \Omega}{4 \pi}
{\text tr} [P^{(a)} \phi'].
\label{Pphiprime}
\end{equation}
For $T=1/2$ case, by using
(\ref{P1phi}) (\ref{P2phi}) for $P^{(a)}$,
we obtain the same result as
(\ref{PTCfor1}) (\ref{PTCfor2}).
For $T \ge 1$,
$P^{(a)}$ is given by (\ref{Pamulti}).
By using (\ref{def_NC_phi}),
taking commutative limit of (\ref{Pamulti}),
and inserting it into (\ref{Pphiprime}),
we can obtain the same result as (\ref{NCindex}).
 
We now consider the commutative limit of the configuration
(\ref{higerisospinai}).
We regard it as a configuration in the $SU(2)$ gauge theory
coupled with the fermion in $2T+1$ dimensional representation,
although we could also regard it as a configuration
in the $SU(2T+1)$ gauge theory coupled with
the fermion in the fundamental representation.
Then we obtain the same commutative limit as we did in
subsection \ref{sec:tpmpconf}.
Eqs.(\ref{TPsolutionai}) (\ref{TPsolution})
imply that the configuration (\ref{higerisospinai})
has the winding number of $1$.
 
We now define another topological charge:
\begin{equation}
\frac{1}{2}{\cal T}r [\phi'(\Gamma^R + \hat\Gamma)],
\label{phiprimeTC}
\end{equation}
where we insert $\phi'$ of (\ref{def_NC_phi})
instead of inserting the projection operator
as in (\ref{proj_index}).
On the fuzzy 2-sphere it is evaluated as
\begin{equation}
\sum_{a=1}^{2T+1}
\frac{L^{(a)}(L^{(a)}+1)-L(L+1)}{2L+1}
2(L-L^{(a)})
= -\frac{2}{3}T(T+1)(2T+1)
\label{ncvalueofphiTC}
\end{equation}
for the configuration (\ref{higerisospinai}).
On the other hand, by taking the weak gauge limit and
the commutative limit, this becomes
\begin{equation}
\frac{1}{2}{\cal T}r [\phi' (\Gamma^R + \hat\Gamma)]
\to \rho^2 \int  \frac{d \Omega}{4 \pi}
{\text tr}\Big[
 \phi'\epsilon_{ijk} n_i F_{jk} \Big],
\label{comlimphiprimeTC}
\end{equation}
and then, by using $SU(2)$ condition
(\ref{SU2com2}), it is calculated as
\begin{equation}
-2 \int  \frac{d \Omega}{4 \pi}
{\text tr} [ \phi'^2]
=-\frac{2}{3}T(T+1)(2T+1),
\end{equation}
which agrees with (\ref{ncvalueofphiTC}).
Also, for the configurations (\ref{puregaugeconfig}),
(\ref{comlimphiprimeTC}) becomes
\begin{equation}
-\frac{2}{3}T(T+1)(2T+1)
\frac{\rho^2}{8\pi}\int_{S^2} d\Omega \epsilon_{ijk}
n_i \epsilon^{abc}
\phi'^a (\partial_j \phi'^b) (\partial_k \phi'^c),
\end{equation}
which is the winding number of
$\pi_2(SU(2)/U(1))$.
Comparing with the above values,
we see that the winding number is $1$
for the configuration (\ref{higerisospinai}),
and for configurations which satisfy
the $SU(2)$ algebra (\ref{SU2alg})
in general.
 
We thus obtain a configuration with winding number $1$,
by suitably combining the coordinates of fuzzy 2-sphere
and the gauge configuration, both of which
satisfy the $SU(2)$ algebra.
It will be an interesting attempt
to construct configurations
with general winding numbers
by generalizing the procedure.
 
\subsection{Other configurations}
 
In this subsection we consider other configurations.
 
First we consider the case where the fermion is in
$m$ dimensional representaion of the gauge group,
for example,
$U(m)$ gauge theories coupled with the fermion
in the fundamental representaion.
Then the Hilbert space on which $A_i$ acts is
$m(2L+1)$ dimensional.
We here assume that $A_i$ satisfy the $SU(2)$ algebra.
Then $A_i$ is decomposed into
irreducible representations as
\begin{equation}
A_i\simeq
\begin{pmatrix}
 L_i^{(1)} & & & \cr
& L_i^{(2)} & & \cr
& & \ddots & \cr
& & & L_i^{(r)} \cr
\end{pmatrix}
\end{equation}
where $L_i^{(a)}$'s are $ L^{(a)}$
representations of $SU(2)$ for $a=1,\cdots,r$.
The projection operator to pick up the Hilbert space
on which $L_i^{(a)}$ acts is
defined to be (\ref{Pamulti}).
Then we have the same topological charge
as (\ref{proj_index})
and the same result as (\ref{NCindex}).
Since
$\sum_{a=1}^r (2L^{(a)}+1)=m(2L+1)$,
we have
\begin{eqnarray}
\frac{1}{2}{\cal T}r (\Gamma + \hat\Gamma)
&=& \frac{1}{2}\sum_{a=1}^{r}
{\cal T}r P^{(a)} (\Gamma + \hat\Gamma) \n \\
&=&\sum_{a=1}^{r} 2(L-L^{(a)}) \n \\
&=&(r-m)(2L+1),
\end{eqnarray}
which means that the topological charge
without the projection operator
gives a multiple number of $2L+1$ .

Next we consider a NC analogue of
$U(1)$ Dirac monopoles.
Let's consider the configuration
\begin{equation}
A_i \simeq
\begin{pmatrix}
 L_i^{(1)} & & & \cr
& c_i^{(1)} & & \cr
& & \ddots & \cr
& & & c_i^{(s)} \cr
\end{pmatrix},
\end{equation}
where $L_i^{(1)}$
is $L^{(1)}$ representation of $SU(2)$ algebra, and 
$c_i^{(a)}$'s $(a=1,\cdots,s)$ are some numbers
($1\times 1$ matrices).
We note here that the $A_i$ does not satisfy the $SU(2)$
algebra, but does satisfy the equation of motion in the
reduced model of 3-dimensional Yang-Mills theory
with the Chern-Simons term\cite{IKTW}:
$\left[ A_i, \left[ A_i, A_j \right] \right] =
-i\epsilon_{jkl} \left[ A_k, A_l \right]$.
The topological charge is calculated as
\begin{equation}
\frac{1}{2}{\cal T}r_{(1)}(\Gamma+\hat\Gamma)=
2(L-L^{(1)})=s .
\end{equation}
We may be able to regard this solution
as ``D2 with $s$ D0's''\cite{Hashimoto:2001xy},
or ``Dirac monopole
with $s$ Dirac strings''\cite{Karabali:2001te}. 
But in the latter interpretation there are several problems 
which should be resolved.

\section{Discussion}
\setcounter{equation}{0}
In this paper, after we briefly review the construction of
GW Dirac operator and chirality operators
on fuzzy 2-sphere \cite{AIN2},
we further investigate several topological properties.
First we calculated the chiral anomaly
from the non-trivial Jacobian
up to the second order in gauge fields
and showed that the correct form of the Chern character
is reproduced in the commutative limit.
This result completed the first order
calculation of the chiral anomaly in \cite{AIN2}.
Thus we saw that the topological charge we defined on the
fuzzy sphere has the correct commutative limit.

We then studied topologically nontrivial configurations.
Even in the theories on the commutative sphere,
if we naively integrate the topological charge density
over the sphere,
the topological charge vanishes identically for
any configurations.
In the 't Hooft-Polyakov (TP) monopole,
we introduce the idea of spontaneous symmetry breaking,
insert the projector to pick up the unbroken $U(1)$ component
into the topological charge,
and obtain nonzero values for it.
These idea and procedure correspond to
introducing some projections into
matrices in the noncommutative (NC) theories.
We considered the TP monopole configurations
and their topological charges on the fuzzy sphere,
and showed that they have the correct commutative limit.
We further studied other nontrivial configurations
and their topological charges.

Projections into matrices may be necessary not only for
describing the topologically nontrivial configurations,
but for describing the smooth gauge configurations themselves,
and the smooth space-time itself.
Since the denominator of $\hat\Gamma$ is given by $\sqrt{H^2}$,
we need to avoid zero eigenvalues of $H$,
which corresponds to the admissibility conditions
in the lattice gauge theories.
Also, since space-time and field-theoretical degrees of freedom
are considered to be
embedded in the near-diagonal elements of the matrices,
we need to project these elements from the full matrices
to describe smooth field-configurations and space-time.
Studies of embeddings of
classical configurations in matrices
\cite{Kiskis:2002gr}\cite{Kikukawa:2002ms}
may be useful for this study.
 
It is also interesting to construct GW fermions
on various NC geometries,
such as NC torus\cite{Nishimura:2001dq,Iso:2002jc}
and NC $CP^{n}$,
and consider the above mentioned problems
in these models.
Also, it may be interesting to study
the Seiberg-Witten map\cite{SW} on fuzzy
2-sphere\cite{Hayasaka:2002db}
to investigate nontrivial configurations
and topological charges
in commutative and NC theories.
 
We expect that
our formalism can provide a clue for classifying the
space-time topology as well as
the topology of gauge field space,
since space-time and
matter are indivisible in matrix models
or NC field theories.
We hope that our formalism based on the GW relation will
have important roles for considering topological structures
of space-time and matter in
these theories.
 
\section*{Acknowledgements}
We would like to thank
U. Carow-Watamura and S. Watamura for useful discussions.
The content of the paper was presented
at the string and field theory workshop at KEK on March
18-20(19), 2003 by S. Iso, and at the conference Lattice 2003
at Tsukuba on July 15-19(16), 2003 by K. Nagao\cite{nagaolat03}.
 
\appendix
 
\section{Proof of Index Theorem}\label{sec:indextheo}
\setcounter{equation}{0}
 
In this appendix we prove the index theorem (\ref{indexth}).
We defined
the chirality operator and the GW Dirac operator
more generally in \cite{AIN2},
so here we prove the index theorem in this general formulation.
It is much easier to prove it in the formulation of this paper.
 
We defined in \cite{AIN2} the GW Dirac operator $D_{\rm GW}$ as
\begin{equation}
f(a,\Gamma) D_{\rm GW}=1- \Gamma \hat{\Gamma},
\label{GWdef2}
\end{equation}
where $\Gamma$ and $\hat\Gamma$ are
the generalization of $\Gamma^R$ and $\hat\Gamma$ of this paper,
which are Hermite, satisfy $\Gamma^2=\hat\Gamma^2=1$,
and become the commutative chirality operator
in the commutative limit.
The prefactor $f(a,\Gamma)$ can be
any function of $a$ and $\Gamma$,
but must be of the order of $a$,
and must be invertible.
Hermite conjugate of (\ref{GWdef2}) gives
\begin{equation}
D_{\rm GW}^\dagger  f(a,\Gamma)^\dag =1- \hat{\Gamma}\Gamma.
\label{GWdefhc}
\end{equation}
By multiplying (\ref{GWdef2}) by $\Gamma$ from the left,
multiplying another (\ref{GWdef2})
by $\hat\Gamma$ from the right, and summing them, we obtain
the following GW relation:
\begin{eqnarray}
\Gamma D_{\rm GW}+ D_{\rm GW} \hat\Gamma =0,
\label{GWrel}\\
D_{\rm GW}^\dagger \Gamma +\hat\Gamma D_{\rm GW}^\dagger=0.
\label{GWrelhc}
\end{eqnarray}
Also, since (\ref{GWdef2}) multiplied  by $\Gamma$ both from
the left and the right becomes (\ref{GWdefhc}),
\begin{eqnarray}
D_{\rm GW}^\dagger&=&\Gamma f(a,\Gamma) D_{\rm GW}
\Gamma f(a,\Gamma)^{\dagger -1},
\label{gammagamma}\\
D_{\rm GW}&=& f(a,\Gamma)^{-1} \Gamma
D_{\rm GW}^\dagger f(a,\Gamma)^\dagger \Gamma.
\label{gammagammahc}
\end{eqnarray}
Similarly, by multiplying (\ref{GWdef2}) by $\hat\Gamma$,
we obtain
\begin{eqnarray}
D_{\rm GW}^\dagger&=&\hat\Gamma f(a,\Gamma) D_{\rm GW}
\hat\Gamma f(a,\Gamma)^{\dagger -1},
\label{gammahatgammahat}\\
D_{\rm GW}&=& f(a,\Gamma)^{-1} \hat\Gamma
D_{\rm GW}^\dagger f(a,\Gamma)^\dagger \hat\Gamma.
\label{gammahatgammahathc}
\end{eqnarray}
 
Now we introduce Fock space ${\cal H}$ for the spinor
matrices $\psi$.
In the case of 2-sphere, ${\cal H}$ is an ensemble of
$(2L+1) \times (2L+1)$ Hermitian matrices.
The above GW Dirac operator $D_{\rm GW}$
and the chirality operators
$\Gamma$, $\hat\Gamma$ act on this space.
We then decompose ${\cal H}$ into the spaces of
zero and nonzero eigenmodes for $D_{\rm GW}$,
and also for $D_{\rm GW}^\dagger$:
\begin{eqnarray}
{\cal H}&=& {\cal H}_0\oplus \bar{{\cal H}_0},\\
&=& {\cal H}'_0\oplus \bar{{\cal H}'_0},
\end{eqnarray}
where
\begin{eqnarray}
{\cal H}_0&=&\{\psi \in {\cal H} | D_{\rm GW} \psi=0 \},\\
{\cal H}_0'&=&\{\psi \in {\cal H} | D_{\rm GW}^\dagger \psi=0 \},
\end{eqnarray}
and $\bar{{\cal H}_0}$, $\bar{{\cal H}'_0}$ are their
complementary space.
 
First we show the following statement:
\begin{eqnarray}
&&{\cal H}_0={\cal H}'_0, \
\bar{{\cal H}_0}=\bar{{\cal H}'_0}, \n \\
&&\psi\in {\cal H}_0={\cal H}'_0 \Rightarrow
\Gamma \psi= \hat\Gamma \psi \in {\cal H}_0={\cal H}'_0.
\label{statement1}
\end{eqnarray}
Its proof is as follows:
If $\psi \in {\cal H}_0$, then due to (\ref{GWrel}),
$\hat\Gamma\psi \in {\cal H}_0$.
(\ref{GWdef2}) leads to $\Gamma\psi=\hat\Gamma\psi$.
Then $\Gamma\psi\in {\cal H}_0$.
Thus, due to (\ref{gammagamma}),
$D_{\rm GW}^\dagger \psi=0$, and then $\psi\in {\cal H}'_0$.
Hence, ${\cal H}_0 \subset {\cal H}'_0$.
Similarly,
if $\psi \in {\cal H}'_0$, then due to (\ref{GWrelhc}),
$\Gamma\psi \in {\cal H}'_0$.
Thus, due to (\ref{gammagammahc}),
$D_{\rm GW}\psi=0$, and then $\psi\in {\cal H}_0$.
Hence, ${\cal H}'_0 \subset {\cal H}_0$.
Therefore ${\cal H}'_0 = {\cal H}_0$.
$\bar{\cal H}'_0 = \bar{\cal H}_0$ is
its contraposition.
 
Next we show
\begin{equation}
\psi\in \bar{\cal H}_0=\bar{\cal H}'_0 \Rightarrow
\Gamma \psi, \hat\Gamma \psi \in \bar{\cal H}_0=\bar{\cal H}'_0.
\label{statement2}
\end{equation}
We prove its contraposition as follows :
If $\Gamma\psi \in {\cal H}'_0$,
$D_{\rm GW}^\dagger \Gamma\psi=0$.
Then, due to (\ref{GWrelhc}),
$\hat\Gamma D_{\rm GW}^\dagger \psi=0$.
Multiplying it by $\hat\Gamma$ from the left we have
$D_{\rm GW}^\dagger \psi=0$.
Thus $\psi\in {\cal H}'_0$.
Similarly, if $\hat\Gamma\psi \in {\cal H}_0$,
$D_{\rm GW} \hat\Gamma\psi=0$.
Then, due to (\ref{GWrel}),
$\Gamma D_{\rm GW} \psi=0$.
Multiplying it by $\Gamma$ from the left we have
$D_{\rm GW} \psi=0$.
Thus $\psi\in {\cal H}_0$.
 
Finally,
\begin{equation}
\psi \in \bar{\cal H}_0 =\bar{\cal H}'_0\Rightarrow
\Gamma\psi, \hat\Gamma\psi  \ {\rm anti-pairing}
\label{statement3}
\end{equation}
since if $\Gamma\psi = \pm \psi$,
then due to (\ref{GWrelhc}),
$\hat\Gamma (D_{\rm GW}^\dagger \psi)
=-D_{\rm GW}^\dagger \Gamma \psi
=\mp (D_{\rm GW}^\dagger \psi)$.
Similarly,
if $\hat\Gamma\psi = \pm \psi$,
then due to (\ref{GWrel}),
$\Gamma (D_{\rm GW} \psi)
=-D_{\rm GW} \hat\Gamma \psi
=\mp (D_{\rm GW} \psi)$.
 
{}From (\ref{statement1}), (\ref{statement2}), (\ref{statement3}),
we can prove the index theorem:
\begin{eqnarray}
{\cal T}r(\Gamma + \hat\Gamma)
&=& {\cal T}r_{{\cal H}_0}(\Gamma + \hat\Gamma)
+{\cal T}r_{\bar{\cal H}_0}(\Gamma + \hat\Gamma)\n \\
&=&{\cal T}r_{{\cal H}_0}(\Gamma + \hat\Gamma)\n \\
&=&2(n_+ -n_-)\n \\
&=&2 \ {\rm Index}(D_{\rm GW}).
\end{eqnarray}

\section{Proof of $\delta \left({\cal T}r \hat\Gamma \right) =0$}
\label{sec:deltrhatgamma}
\setcounter{equation}{0}
In this appendix we show that
if we define $\hat\Gamma$ as in (\ref{gammahat}),
${\cal T}r (\hat{\Gamma})$ is invariant
under any small deformation of any parameter or any
configuration such as gauge field in the operator $H$.
Under the infinitesimal deformation of $H$,
$H\rightarrow H+\delta H$,
${\cal T}r (\hat{\Gamma})$ varies as
\begin{eqnarray}
\delta \left( {\cal T}r \hat{\Gamma} \right)
&=&\delta \left({\cal T}r H \frac{1}{\sqrt {H^2}} \right)
\nonumber \\
&=& {\cal T}r\left( \delta H \frac{1}{\sqrt {H^2}} \right)
+{\cal T}r
\left(H \frac{1}{\sqrt {(H+\delta H)^2}} \right)
- {\cal T}r \left(H \frac{1}{\sqrt {H^2}} \right). \n
\end{eqnarray}
The second term can be evaluated as
\begin{eqnarray}
&&{\cal T}r\left(H\frac{1}{\sqrt{(H+\delta H)^2}}
\right)\nonumber \\
&=&
{\cal T}r\left(
H \int_{-\infty}^{\infty} \frac{dt}{\pi}
\frac{1}{t^2+H^2+H\delta H + (\delta H) H + (\delta H)^2}
\right) \nonumber \\
&=&
{\cal T}r\left[ H \int_{-\infty}^{\infty} \frac{dt}{\pi}
\frac{1}{t^2 +H^2} \right] \nonumber \\
&&-{\cal T}r\left[H \int_{-\infty}^{\infty} \frac{dt}{\pi}
\frac{1}{t^2+H^2}
(H\delta H +\delta H H) \frac{1}{t^2+H^2} \right]
+{\cal O}((\delta H)^2)
\nonumber \\
&=&
{\cal T}r\left[ H \frac{1}{\sqrt {H^2}} \right]
-{\cal T}r\left[ \frac{1}{\sqrt {H^2}} \delta H \right]
+{\cal O}((\delta H)^2) ,
\end{eqnarray}
where we have utilized the cyclic property in
${\cal T}r$ and the following identities,
\begin{eqnarray}
\frac{1}{\sqrt {X^2}}&=&
\int_{-\infty}^{\infty} \frac{dt}{\pi} \frac{1}{t^2 +X^2}, \\
\frac{1}{2 X^2 \sqrt{X^2}}&=&
\int_{-\infty}^{\infty} \frac{dt}{\pi} \frac{1}{(t^2 +X^2)^2}.
\end{eqnarray}
Therefore we obtain
\begin{equation}
\delta \left({\cal T}r \hat\Gamma \right)= 0.
\end{equation}

\end{document}